\documentclass[twocolumn,aps,pra,nofootinbib]{revtex4-1}

\usepackage{graphicx} 
\usepackage{bm}
\usepackage{natbib}
\usepackage{amsmath}

\newcommand{\eref}[1]{Eq.~(\ref{#1})}
\newcommand{\tref}[1]{Table~\ref{#1}}

\begin{document}
\preprint{APS/123-QED}

\title{Low-lying energy levels of $^{229}$Th$^{35+}$ and the electronic bridge process}

\author{S. G. Porsev$^{1,2}$, C. Cheung$^{1}$, and M. S. Safronova$^{1}$}

\affiliation{
$^{1}$Department of Physics and Astronomy, University of Delaware, Newark, Delaware 19716, USA\\
$^{2}$Petersburg Nuclear Physics Institute of NRC ``Kurchatov Institute'', Gatchina, Leningrad district 188300, Russia
}

\begin{abstract}
The nuclear transition between the ground and the low-energy isomeric state in the $^{229}$Th nucleus is of
interest due to its high sensitivity to a hypothetical temporal variation of the fundamental constants and
a possibility to build a very precise nuclear clock, but precise knowledge of the nuclear clock transition frequency
is required. In this work we estimate the probability of an electronic bridge process in $^{229}$Th$^{35+}$, allowing
to determine the nuclear transition frequency and reduce its uncertainty. Using configuration interaction
methods we calculated energies of the low-lying states of Th$^{35+}$ and determined their uncertainties.
\end{abstract}

\maketitle
\section{Introduction}
A specific feature of the $^{229}$Th nucleus is that the energy difference between the ground state and the first excited state
is only several eV. An existence of such a low-lying level was established more than forty years ago~\cite{KroRei76},
but a precise measurement of the isomeric state energy turned out to be very difficult.
In 1994 Helmer and Reich~\cite{HelRei94} measured this excitation energy ($\omega_N$) to be $3.5 \pm 1$ eV.
In Ref.~\cite{GuiHel05} it was obtained as $5.5 \pm 1.0$ eV.
The experiments of Beck {\it et al.}~\cite{BecBecBei07,BecWuBei09} gave an even larger value with less error,
$\omega_N = 7.8 \pm 0.5$ eV. Finally, in recent experiments of Seiferle {\it et al.}~\cite{SeiWenBil19}
and Sikorsky {\it et al.}~\cite{SikGeiHen20}
the values $\omega_N = 8.28 \pm 0.17$ and $\omega_N = 8.10 \pm 0.17$ eV were obtained. Both these results were used in an analysis
carried out in Ref.~\cite{PeiSchSaf21} yielding the value $\omega_N = 8.19 \pm 0.12$ eV.
Thus, the current most precise value differs from the result of 1994 by more than two times,
but is in a good agreement with the result obtained in Ref.~\cite{BecWuBei09}.

The nuclear transition between the ground and the low-energy isomeric state in the $^{229}$Th nucleus is of
interest due to its high sensitivity to a hypothetical temporal variation of the fundamental constants \cite{PeiTam03}.
Another unique feature of this transition is a possibility to build a very precise nuclear clock~\cite{Fla06}.
It requires the precise knowledge of the nuclear clock transition frequency and, consequently, further investigations aiming
to refine the value of $\omega_N$ are needed.

An experimental progress in trapping and sympathetic cooling of highly charged ions (HCIs) (see review in Ref.~\cite{KozSafCre18} for details)
using electron-beam ion traps (EBITs) opened new possibilities to use optical transitions of such ions for different applications.
As it was discussed in Ref.~\cite{KozSafCre18}, an interaction region of an electron beam with maximum magnetic field in EBITs is short enough
that reduces possible electron-beam instabilities and also allows for a high charge density. Both of these effects speed up the ionization process.
If a highly charged ion, considered as a clock candidate, does not have a transition suitable for
laser cooling, sympathetic cooling can be done by trapping the HCI together with the
cooling ion in the same trapping potential and using their mutual Coulomb interaction.
In the recent paper~\cite{BilBekBer20} it was suggested to use the electronic
bridge (EB) process in Th$^{35+}$ for an accurate determination of the nuclear isomeric energy and the energies of the low-lying
states, needed to estimate the EB process rate, were calculated.

Since the EB process rate depends drastically on the energies of the states involved in this process,
we calculated the energies of the low-lying states of Th$^{35+}$ in different approximations of increasing complexity.
The main configuration of the ground state of the ion is $(1s^2 ... 4d^{10} 4f^9)$ and we used configuration interaction (CI) method for calculations.
First, we performed a 9-electron CI calculation, including nine $4f$ electrons in the valence field while treating all other
electrons as the core electrons. Then, we carried out 19- and 25-electron CI calculations, including the $4f,4d$ and $4f,4d,4p$ electrons
into the valence field, respectively. In such a way we were able to estimate the role of core-valence correlations, what allowed us to
determine the uncertainties of the energies.

To carry out these calculations for such a complicated open-shell system as Th$^{35+}$
we used a new parallel atomic structure code package developed and described in Ref.~\cite{CheSafPor21}.
This package (i) allows us much quicker computations and
(ii) enables to work with a CI space of a much larger size than was possible previously when we used
serial versions of the programs. Using a parallel version
of the program for the CI calculation, we were able to use several hundreds of processors simultaneously and, consequently, to work
with a large CI space, involving up to 120 millions of determinants.

Using the energies obtained and assuming a resonant character of the induced EB process, we estimated its rate for several possible values
of the nuclear isomeric energy. Based on these calculations we conclude that modern facilities of EBITs and available
ultra-violet lasers allow us to determine more precisely the nuclear isomeric energy and reduce its uncertainty.
\section{Electronic bridge excitation}
\label{EB}
\begin{figure}[h]
 \includegraphics[scale=0.5]{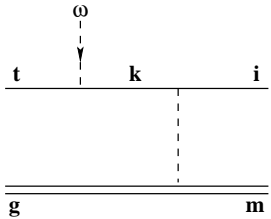}
 \caption{The Feynman diagram of the considered electronic bridge process. The single and double solid lines relate to the electronic
and the nuclear transitions, correspondingly. The dashed line is the photon line.}
 \label{Fig:EB2}
 \end{figure}

Here we consider the process of the excitation of the nucleus from the ground ($g$) to the isomeric state ($m$) by an
electronic bridge process driven by a one-photon excitation of the electron shell. Such a
process, relying on the absorption of an incident photon, can be represented by the Feynman diagram in Fig.~\ref{Fig:EB2}.
In this process the electronic shell is promoted by a laser photon from its initial state $t$ to an excited state $k$ and
then decays to a lower-lying state $i$. The energy transferred to the nucleus is used to excite it from the ground to isomeric state.
Assuming a resonant character of the process, we take into account only one Feynman diagram that gives the main contribution
to the probability of the process.

In the following we use the formalism developed in Refs.~\cite{PorFlaPei10,PorFla10ThIV,PorFla10ThII}
with the difference that an $M1$ photon (instead of a $E1$ photon) is absorbed by the electron shell.
Such a photon is described by the magnetic moment operator
\begin{equation}
{\bm \mu} = -\mu_0\, ({\bf J} + {\bf S}) .
\end{equation}
Here ${\bf J}$ and ${\bf S}$ are the total and spin momentum operators and $\mu_0$ is the Bohr magneton
determined as $\mu_0 = |e| \hbar/(2mc)$, where $e$ and $m$ is the electron charge and mass, $c$ is the speed of light,
and $\hbar$ is the Plank constant.
(If not stated otherwise the atomic units $\hbar = m = |e| = 1$ and $c \approx 137$ are used in the following).

We assume that the incident radiation with spectral intensity $I_\omega$ is isotropic and unpolarized.
Following Ref.~\cite{Sob79} the relation between the EB excitation rate, $\Gamma_{\rm exc}$, and the rate of the inverse spontaneous
EB process, $\Gamma$, that can be formally described by the mirror-image of Fig.~\ref{Fig:EB2} with an outgoing photon arrow,
can be written as
\begin{equation}
\Gamma_{\rm exc} = \Gamma \,  \frac{4 \pi^3 c^2}{\omega^3} g_r I_\omega ,
\label{Wa}
\end{equation}
where $\omega$ is the frequency of the absorbed photon. The factor $g_r \equiv (2I_m+1)(2J_i+1)/[(2I_g+1)(2J_t+1)]$
appears because we average over the initial projections and summing over the final projections of the electronic
and nuclear total angular momenta. Here $I_m$ ($I_g$) is the nuclear spin of the isomeric (ground) state
and $J_i$ ($J_t$) is the total angular momentum of the initial (final) electronic state in the spontaneous EB process.

The general expression for the probability of the EB process was derived in Ref.~\cite{PorFla10ThIV}.
Assuming a resonance character of the spontaneous EB process, its transition rate can be written as,
\begin{eqnarray}
\Gamma \approx \frac{16\, \pi}{3\,(2J_i+1)} \left(\frac{\omega}{c}\right)^3
\sum_{K=1}^2 \frac{B(\tau K)\, G_K}{(2K+1)^2} .
\label{GamK}
\end{eqnarray}

The reduced probabilities, $B(M1)$ and $B(E2)$, of the nuclear magnetic-dipole and electric-quadrupole
$m \rightarrow g$ transition are determined as
\begin{equation}
B(\tau K) = \frac{1}{2I_m+1} \frac{2K+1}{4\pi} |\langle I_g || \mathcal{M}_K || I_m \rangle|^2 ,
\label{B}
\end{equation}
where $\tau \equiv M$ for $K=1$ and $\tau \equiv E$ for $K=2$;
$\mathcal{M}_1$ and $\mathcal{M}_2$ are the magnetic-dipole and electric-quadrupole nuclear moment operators,
and $|I_g \rangle = 5/2^+$ [633]  and $|I_m \rangle = 3/2^+$ [631] are the ground
and isomeric nuclear states, respectively, given in their Nilsson classification.


The explicit expression for the quantity $G_K$ was derived in Ref.~\cite{PorFla10ThIV} and in our case is reduced to
\begin{eqnarray}
G_K \approx \frac{1}{2J_k+1} \left\vert
\frac{\langle i ||T_K||  k \rangle \langle k ||\mu|| t \rangle}{E_k - E_i -\omega_N} \right\vert^2 ,
\label{G2}
\end{eqnarray}
where $|i\rangle$, $|k\rangle$, and $|t\rangle$ are the electronic states, $E_k$ and $E_i$ are the energies of the respective states,
$J_k$ is the electron total angular momentum of the state $|k\rangle$,
and $\omega_N$ is the nuclear transition frequency.
We assume that there is only one state giving the dominant contribution to $G_K$, while the contribution of all other intermediate states
is negligible, i.e., the state $|k\rangle$ is chosen so that $E_k - E_i \approx \omega_N$.

The expressions for the single-electron operators $T_1$ and $T_2$ can be written as
\begin{eqnarray}
T_{1 \lambda}({\bf r}) &=& -\frac{i \sqrt{2}\, {\bm \alpha} \cdot {\bf C}^{(0)}_{1 \lambda}({\bf n})}{c\,r^2} , \nonumber \\
T_{2 \lambda}({\bf r}) &=& -\frac{C_{2 \lambda}({\bf n})}{r^3} ,
\end{eqnarray}
where ${\bm \alpha}$ is the Dirac matrix, ${\bf n} \equiv {\bf r}/r$, and ${\bf C}^{(0)}_{K \lambda}$ is a normalized vector spherical harmonic
defined by (see, e.g., Ref.~\cite{VarMosKhe88})
\begin{equation}
{\bf C}^{(0)}_{K \lambda}({\bf n}) = \frac{{\bf L}} {\sqrt{K(K+1)}} C_{K \lambda}({\bf n}) .
\end{equation}
Here ${\bf L}$ is the orbital angular-momentum operator and
$C_{K \lambda}$ is a spherical harmonic given by
\begin{equation}
 C_{K \lambda}({\bf n}) = \sqrt{\frac{4\pi}{2K+1}}\, Y_{K \lambda}({\bf n}) .
\end{equation}

A most accurate determination of the nuclear transition frequency $\omega_N$ was recently done in Ref.~\cite{PeiSchSaf21} to be
$\omega_N = 8.19(12)\, {\rm eV}$. The authors of Ref.~\cite{BilBekBer20} calculated the low-lying energy levels of Th$^{35+}$
and found the transition frequency from the low-lying excited state with $J=13/2$ to the ground state ($J=15/2$) to be 8.40 eV
which is close to $\omega_N$. The state with $J=11/2$ and $E \approx 4.2 \, {\rm eV}$ can be used as the initial state in the induced
EB process. Applying the designations in Fig.~\ref{Fig:EB2} we have $|t \rangle \equiv |J=11/2\rangle$, $|k \rangle \equiv |J=13/2\rangle$,
and $|i \rangle \equiv |J=15/2\rangle$.

We carried out the calculation of the low-lying states in the framework of the CI method
using the program package described in Ref.~\cite{KozPorSaf15} and further developed in Ref.~\cite{CheSafPor21}.
We included the Breit interaction correction and estimated the core-valence correlations.
 Our results, discussed more detailed in the following sections,
confirmed the main conclusion of Ref.~\cite{BilBekBer20} about a possibility to determine accurately the nuclear isomer energy using the EB process
with realistic laser parameters.
\section{Method of calculation and results.}
We start calculations of energy levels of Th$^{35+}$ belonging to the $(1s^2,...,4f^9)$ configuration
from solution of the Dirac-Hartree-Fock (DHF) equations, carrying out a self-consistency procedure for the $[1s^2,...,4f^9]$ electrons.
The Breit interaction was included in this procedure.
The remaining virtual orbitals were formed using a recurrent procedure described in~\cite{KozPorFla96,KozPorSaf15}, when the large component
of the radial Dirac bispinor, $f_{n'l'j'}$, was obtained from a previously constructed function $f_{nlj}$ by
multiplying it to $r^{l' - l}\, \sin(kr)$, where $l'$ and $l$ are the orbital quantum numbers of the new and old orbitals ($l' \geq l$) and the coefficient $k$ is determined by the properties of the radial grid. The small component $g_{n'l'j'}$ was found from the kinetic balance condition:
\begin{equation}
\label{kbal}
g_{n'l'j'} =\frac{\bm \sigma \bm p}{2mc} f_{n'l'j'} ,
\end{equation}
where $\bm\sigma$ are the Pauli matrices, ${\bm p}$ and $m$ are the electron momentum and mass, and $c$ is the speed of light.
The newly constructed functions were then orthonormalized with respect to the functions of the same symmetry.
In constructing the virtual orbitals we did not diagonalize the basis set.

In total the basis set consisted of six partial waves ($l \leq 5$) including the orbitals up to $10s$, $10p$, $10d$, $10f$, $10g$, and $10h$.
The configuration space grows very rapidly with an increase of the basis set; for this reason the basis set is rather short.

For an accurate calculation of energies we need to take into account valence-valence and core-valence correlations. The former can be
treated explicitly in the framework of the CI method. A CI many-electron wave function of a given angular momentum J and parity
can be represented by a linear combination of Slater determinants~\cite{DzuFlaKoz96}:
\begin{equation}
\Psi_J = \sum_k a_k \Phi_k .
\end{equation}

A way to account for core-valence correlations was suggested in Ref.~\cite{DzuFlaKoz96}, but in practice it is applicable when
the number of valence electrons does not exceed 4-5. In our consideration we include nine $4f$ electrons into the valence field and
such a method is impractical. To estimate the role of different core shells, we carried out several calculations of increasing complexity
in the framework of the CI method sequentially adding the core shells into the valence field.
In this way we carried out the 9-, 19- and 25-electron CI calculations when the $4f$, ($4f,4d$), and ($4f,4d,4p$),
electrons, respectively, were included in the CI space. Below we discuss these calculations more detailed.
\subsection{9-electron CI}
We start from the most simple case of the 9-electron CI.
To check the convergence of the CI method, we calculated the low-lying energy levels for five cases. In the first case we included the
single and double excitations of the electrons from the main configuration $4f^9$ to the $6s$, $6p$, $6d$, $6f$, $6g$ and $6h$ orbitals
(we designate it as [$6spdfgh$]). In other four calculations, we included the single and double excitations to [$n\,spdfgh$], where $n=7-10$.
We checked that an inclusion of the triple excitations change the energies only by few cm$^{-1}$. Thus, this contribution can be neglected.

The results are presented in \tref{Energ}. The terms are determined by their total angular momentum $J$. The energies of four lowest-lying excited states counted from the ground state and
found for different $n$ (when the single and double excitations were allowed to [$n\,spdfgh$]) are presented in the columns
labeled ``$E$'' in cm$^{-1}$. The energy
differences $\Delta(n) \equiv E(n\,spdf\!gh) - E(n-1\,spdf\!gh)$ and the ratios $\delta_n \equiv \Delta(n)/\Delta(n-1)$
are given in the columns labeled ``$\Delta(n)$'' (in cm$^{-1}$) and ``$\delta_n$''.
The total values are given in the last row.
\begin{table*}[t]
\caption{9-electron CI. The energies of four lowest-lying excited states counted from the ground state and
found for different $n$ (when the single and double excitations were allowed to [$n\,spdfgh$]) are presented in the columns
labeled ``$E$'' in cm$^{-1}$. The energy differences $E(n\,spdf\!gh) - E(n-1\,spdf\!gh)$ are given in the columns labeled ``$\Delta(n)$'' (in cm$^{-1}$).
The ratios $\Delta(n)/\Delta(n-1)$ are listed in the columns labeled ``$\delta_n$''.
The total values, obtained as $E(n=10)+\Delta(n>10)$, are given in the last row.}
\label{Energ}%
\begin{ruledtabular}
\begin{tabular}{ccccccccccccc}
                &\multicolumn{3}{c}{$J\!=\! 9/2$}
                                                 &\multicolumn{3}{c}{$J\!=\!11/2$}
                                                                                 &\multicolumn{3}{c}{$J\!=\! 3/2$}
                                                                                                          &\multicolumn{3}{c}{$J\!=\!13/2$}\\
                &  $E$  &$\Delta(n)$&$\delta_n$  & $E$ &$\Delta(n)$&$\delta_n$    & $E$ &$\Delta(n)$&$\delta_n$ &   $E$     &$\Delta(n)$&$\delta_n$ \\
\hline\\[-0.5pc]
$n = 6$         &  30503    &           &        &  36728    &           &        &  55006    &        &        &   67200   &        &          \\[0.2pc]
$n = 7$         &  30279    &   -224    &        &  36411    &  -316     &        &  54441    &  -565  &        &   67255   &  54.6  &          \\[0.2pc]
$n = 8$         &  30171    &   -108    & 0.483  &  36266    &  -145     & 0.459  &  54175    &  -266  & 0.471  &   67290   &  35.6  &  0.652  \\[0.2pc]
$n = 9$         &  30118    &    -52.8  & 0.487  &  36205    &   -61.2   & 0.421  &  54061    &  -114  & 0.430  &   67313   &  23.1  &  0.648   \\[0.2pc]
$n =10$         &  30090    &    -28.0  & 0.529  &  36178    &   -26.4   & 0.432  &  54010    &   -51.1& 0.447  &   67329   &  15.9  &  0.689   \\[0.2pc]
$q^{\rm a}$     &           &           & 0.500  &           &           & 0.437  &           &        & 0.449  &           &        &  0.663   \\[0.5pc]
$n >10^{\rm b}$ &           &    -35    &        &           &   -13     &        &           &   -29  &        &           &   33   &          \\[0.5pc]
 Total          &$30055    $&           &        &$36165    $&           &        &$53980    $&        &        &$67362    $&        &          \\[0.2pc]
\end{tabular}
\end{ruledtabular}
\begin{flushleft}
$^{a}$See explanation in the text; \\
$^{b}$The contribution to the energies from the configurations containing shells with $n > 10$.
\end{flushleft}

\end{table*}


To estimate the contribution to the energies from the configurations containing shells with $n > 10$, we note that $\delta_n$ for $n=8,9,10$
are numerically close to each other for all terms. Assuming that the same is trues and for
$n >10$ we are able to use the formula for a geometric progression to estimate the respective contribution.

Putting the first term of the geometric progression to be $b_1 \equiv \Delta(7)$, determining $q$ as an average over $\delta_n$ for $n=8,9,10$,
i.e., $q \equiv (\delta_8 + \delta_9 + \delta_{10})/3$, and using the equation for the sum of $k$ terms in the geometric progression,
\begin{equation}
 S_k = \frac{b_1\,(1-q^k)}{1-q},
\end{equation}
we are able to calculate $S_k$ for any $k$. Then we can estimate $\Delta (n>10)$, as
\begin{equation}
 \Delta (n>10) \approx S_k - (\Delta_8+\Delta_9+\Delta_{10}).
\end{equation}

For instance, for the $J\!=\! 9/2$ term we have $b_1 = -224 \, {\rm cm}^{-1}$, $q=0.50$, and
putting $k$ to be equal to 100, we find $S_{100} \approx -448\,\, {\rm cm}^{-1}$ and $\Delta (n>10) \approx -35 \, {\rm cm}^{-1}$.
Carrying out similar calculations for other term, we find the values listed in the row labeled ``$n >10$'' of \tref{Energ}.
The total values, obtained as $E(n=10)+\Delta(n>10)$, are given in the last row of the table.
\subsection{Core-valence correlations}
Due to the large number of valence electrons we were unable to apply the method combining the CI with
a many-body perturbation theory over residual Coulomb interaction~\cite{DzuFlaKoz96} or with linearized coupled-cluster method~\cite{SafKozJoh09}
to find the core-valence correlations. To estimate this contribution to the energies,
we additionally performed the 19- and 25-electron CI calculations, including the $4d$ and $4d,4p$ electrons into the valence field.
In both cases we allowed the single and double excitations of the electrons from the all valence shells
of the main configuration to [$7spdfgh$]. For example, for the $4d^{10}4f^9$ main configuration the single and double
excitations were allowed from the $4d$ and $4f$ shells. We restricted ourselves by the calculation for [$7spdfgh$] because
the set of configurations is sufficiently complete in this case and, on the other hand, such a calculation is not extremely demanding to
computer resources even for the case of the 25-electron CI.

We focus on computing the energies of the $J=11/2$ and $J=13/2$ states that are most important
for an accurate calculation of the probability of the EB process, as we discussed in Sec.~\ref{EB}. Additionally,
if we are interested in only in the states with a large $J$, it facilitates the CI calculation because it allows us to deal with
the CI space of a smaller size. The calculation results of the transition energies obtained for these states
in the framework of 9-, 19-, and 25-electron CI calculations with the excitations allowed to [$7spdfg$] are presented in~\tref{E:CI}.
The main configurations of the valence electrons are given
in the first column. The energies of the excited states, counted from the ground state energy,
are given in the columns labeled ``$E$'' (in cm$^{-1}$). The differences
$\Delta (4d^{10}4f^9) \equiv E(4d^{10}4f^9) - E(4f^9)$ and $\Delta (4p^6 4d^{10}4f^9) \equiv E(4p^6 4d^{10}4f^9) - E(4d^{10} 4f^9)$
are given for each state in the column labeled ``$\Delta$'' in cm$^{-1}$.  The sum $\Delta (4d^{10}4f^9) + \Delta (4p^6 4d^{10}4f^9)$
(we designate it as $\Delta_{\rm Total}$) is presented in the row labeled ``Total''.
\begin{table}[t]
\caption{The energies obtained in the framework of 9-, 19-, and 25-electron CI calculations
with the excitations allowed to [$7spdfgh$], are presented. The main configurations of the valence electrons are given
in the first column. The energies of the excited states, counted from the ground state energy,
are given in the columns labeled ``$E$'' (in cm$^{-1}$). The differences $E(4d^{10}4f^9) - E(4f^9)$ and $E(4p^6 4d^{10}4f^9) - E(4d^{10} 4f^9)$
are given for each state in the column labeled ``$\Delta$'' in cm$^{-1}$. The sum $\Delta (4d^{10}4f^9) + \Delta (4p^6 4d^{10}4f^9)$
is presented in the row labeled ``Total''.}
\label{E:CI}%
\begin{ruledtabular}
\begin{tabular}{lccccc}
                        &  \multicolumn{2}{c}{$J\!=\!11/2$} & \multicolumn{2}{c}{$J\!=\!13/2$} \\
                        &      $E$         &   $\Delta$     &       $E$       &    $\Delta$    \\
\hline \\ [-0.5pc]
$4f^9$                  &     36411        &                &      67255      &            \\[0.2pc]
$4d^{10}4f^9$           &     37400        &      989       &      66948      &     -307   \\[0.2pc]
$4p^6 4d^{10}4f^9$      &     37487        &       87       &      66899      &      -49   \\[0.2pc]
 Total                  &                  &     1076       &                 &     -356   \\[0.2pc]
\end{tabular}
\end{ruledtabular}
\end{table}


Using the values of the energies from Tables~\ref{Energ} and \ref{E:CI} we are able to determine the final value of the energies
for the $J\!=\!11/2$ and $J\!=\!13/2$ terms as the sum of the ``Total'' value given in \tref{Energ} and $\Delta_{\rm Total}$ from \tref{E:CI}.
Thus, we finally obtain $E_{J\!=\!11/2} \approx 37240\, (980)\,\, {\rm cm}^{-1}\, \approx 4.62(12)\, {\rm eV}$ and
$E_{J\!=\!13/2} \approx 67000\, (315)\,\, {\rm cm}^{-1}\, \approx 8.307(39)\, {\rm eV}$.

The uncertainties are mostly determined by the core-valence correlations not taken into account in our consideration. To estimate them we note
that $\Delta (4d^{10}4f^9)$ is an order of magnitude smaller than $\Delta (4p^6 4d^{10}4f^9)$. We assume that a possible contribution from
the $[1s-4s]$ core shells to the energies does not exceed $\Delta (4d^{10}4f^9)$ and we estimate the uncertainty of the energy
as $|\Delta (4d^{10}4f^9)|$.

Our final values for $E_{J\!=\!11/2}$ and $E_{J\!=\!13/2}$ can be compared with the results obtained in Ref.~\cite{BilBekBer20}
to be 4.19 and 8.40 eV, respectively. The difference can be attributed to the Breit correction, omitted in Ref.~\cite{BilBekBer20},
and a more complete inclusion of the valence-valence and core-valence correlations. In particular, when including the $4d$ shell
into the valence field, the authors of Ref.~\cite{BilBekBer20} were obliged to reduce the configuration space (due to limitations
of the computing facilities) disregarding certain double-electron excitations from the $4d$ shell to the virtual orbitals.
\subsection{EB process excitation rate}
\label{EBprob}
Using Eqs.~(\ref{Wa}-\ref{G2}) we are able to estimate the EB process probability. To find the probability of the spontaneous process
given by \eref{GamK} we need to calculate the matrix elements (MEs) of the operators $T_1$, $T_2$, and $\mu$ in~\eref{G2}.

We carried out this calculation for the largest 25-electron CI when single and double excitations were allowed to [$7spdfgh$] and
obtained
\begin{eqnarray}
|\langle J=15/2 ||T_1|| J=13/2 \rangle| &\approx& 2.0\,\, {\rm a.u.} , \nonumber \\
|\langle J=15/2 ||T_2|| J=13/2 \rangle| &\approx& 39\,\, {\rm a.u.} ,  \nonumber \\
|\langle J=13/2 ||\mu|| J=11/2 \rangle| &\approx& 2.28\, \mu_0 .
\end{eqnarray}
The calculation performed in the framework of the 9-electron CI with allowing single and double excitations to [$10spdfgh$] led to the
values that differ by less than 1\% from the results given above.

The values of the reduced probabilities, $B(M1)$ and $B(E2)$, of the nuclear $m \rightarrow g$ transition available in
the literature are contradictory. In Ref.~\cite{DykTka98} the value of $B(M1)$ was found to be 0.048 Weisskopf units (W.u.).
The calculation of Ruchowska {\it et al.}~\cite{RucPloZyl06} led to the value 0.014 W.u. while the recent model calculation
of Minkov and P\'{a}lffy~\cite{MinPal19} predicted the $B(M1)$ value in the limits of 0.005--0.008 W.u., an order of magnitude smaller than
the result reported in Ref.~\cite{DykTka98}.

A similar situation arises for $B(E2)$. Strizhov and Tkalya~\cite{StrTka91}, referring
to Ref.~\cite{BemMcGPor88}, cited the value of several W.u., while in Ref.~\cite{MinPal19} it was found an order of magnitude larger value,
in the limits 29--43 W.u.. In this work we use for an estimate of the EB transition rate
the recent values, $B(M1)= 0.005\, {\rm W.u.} \approx 3.5 \times 10^{-14}\, {\rm a.u.}$ and
$B(E2)= 29\, {\rm W.u.} \approx 3.1 \times 10^{-16}\, {\rm a.u.}$, obtained in Ref.~\cite{MinPal19}.

As follows from Eqs.~(\ref{GamK}-\ref{G2}) the EB transition rate depends substantially from the magnitude of the nuclear transition
frequency $\omega_N$ determined as $8.19(12) \,{\rm eV}$ in Ref.~\cite{PeiSchSaf21}. To illustrate it we calculated these rates for three
values of $\omega_N$: its central value 8.19 eV and two edge values of the uncertainty interval 8.31 and 8.07 eV. We note that the frequency of the absorbed photon $\omega$ is determined as $\omega = \omega_N - E_{J\!=\!11/2}$ and assume that the time-averaged spectral intensity is
$I_\omega \simeq 10^{-3}\, ({\rm W}/{\rm m}^2)\, {\rm s}$ \cite{PorFlaPei10,BilBekBer20}.

The results are presented in \tref{EB}. The possible values of the nuclear transition frequency $\omega_N$ are listed in the first column.
The quantities $G_1$, $G_2$, and $\Gamma$, given by Eqs.~(\ref{G2}) and (\ref{GamK}), are listed in columns 2-4.
We note that for the used values of $B(M1)$ and $B(E2)$, both terms in Eq.~(\ref{GamK}) give approximately the same contribution.
The values of the EB excitation rate per ion, $\Gamma_{\rm exc}$, are presented in the fifth column for different $\omega_N$.

\begin{table}[t]
\caption{Possible values of the nuclear transition frequency $\omega_N$ are listed in the first column.
The quantities $G_1$, $G_2$, and $\Gamma$ are given by Eqs.~(\ref{G2}) and (\ref{GamK}).
The values of the EB excitation rate, $\Gamma_{\rm exc}$, are presented in the forth column.}
\label{EB}%
\begin{ruledtabular}
\begin{tabular}{ccccc}
    $\omega_N ({\rm eV})$ & $G_1\, ({\rm a.u.})$  & $G_2\, ({\rm a.u.})$  & $\Gamma \,({\rm s}^{-1})$ & $\Gamma_{\rm exc} ({\rm s}^{-1})$  \\
\hline \\ [-0.5pc]
       8.31               &     1570              & $5.93 \times 10^5$    & $5.7 \times 10^{-4}$      &      0.3                        \\[0.2pc]
       8.19               &      1.1              &      404              & $3.5 \times 10^{-7}$      & $2.1 \times 10^{-4}$           \\[0.2pc]
       8.07               &      0.3              &       99              & $7.7 \times 10^{-8}$      & $5.0 \times 10^{-5}$
\end{tabular}
\end{ruledtabular}
\end{table}

As seen from \tref{EB}, the largest $\Gamma_{\rm exc}$, obtained in the case of $\omega_N = 8.31\, {\rm eV}$, reaches 0.3 s$^{-1}$. This is due to that
this value of $\omega_N$ is very close to $E_{J\!=\!13/2} \approx 8.307\, {\rm eV}$ leading to a very small denominator in Eq.~(\ref{G2}) and,
respectively, very large $G_{1,2}$ and $\Gamma_{\rm exc}$. For the other two considered values of the nuclear transition frequency, 8.19 and 8.07 eV,
there is no such a resonant enhancement of the effect and the EB excitation rates are several orders of magnitude smaller.
Our results are in a reasonable agreement with those obtained in Ref.~\cite{BilBekBer20}.
\section{Conclusion}
\label{Concl}
We carried out the calculation of the low-lying energy levels for such a complicated multivalent ion as Th$^{35+}$.
To determine the contribution of the valence-valence and core-valence correlations, we performed the 9-, 19-, and 25-electron
CI calculations, including $4f$, $4f,4d$, and $4f,4d,4p$ shells, respectively, into the valence field. Our calculation showed
that the transition energy from the $J=15/2$ state to the ground state, 8.31 eV, is close to the central value of the experimentally
determined nuclear isomer energy, 8.19 eV, and practically coincides with the upper edge value, 8.31 eV. It opens new possibilities for a more
precise measurement of the nuclear isomer energy using an electronic bridge process.

We studied a EB process scheme and estimated the excitation rates of the spontaneous and (inverse) induced EB processes
for possible values of the $g \rightarrow m$ nuclear transition frequency $\omega_N$.
We found the EB excitation rate per ion to be $0.3\,\, {\rm s}^{-1}$ in the case of $\omega_N = 8.31$ eV. For other considered values of
the nuclear transition frequency, 8.19 and 8.07 eV, this rate is 3-4 orders of magnitude smaller.
Based on these results and on the study of Ref.~\cite{BilBekBer20} where typical
electron beam ion trap conditions were considered, we conclude that an efficient population of the nuclear isomer state and
a precise determination of its energy using an EBIT and available ultra-violet lasers is already attainable. The Th$^{35+}$ ion
is a very promising candidate for such an experiment.

We are grateful to P. Bilous and A. P\'{a}lffy for valuable discussion and useful remarks.
This work is a part of the ``Thorium Nuclear Clock'' project that  has received funding from the European Research Council  (ERC) under
the European Union's Horizon 2020 research and innovation program (Grant Agreement No. 856415).
S.P. acknowledges support by the Russian Science Foundation under Grant No. 19-12-00157. This research was supported in part through the use of
the Caviness community cluster at the University of Delaware.


\begin{thebibliography}{27}%
\makeatletter
\providecommand \@ifxundefined [1]{%
 \@ifx{#1\undefined}
}%
\providecommand \@ifnum [1]{%
 \ifnum #1\expandafter \@firstoftwo
 \else \expandafter \@secondoftwo
 \fi
}%
\providecommand \@ifx [1]{%
 \ifx #1\expandafter \@firstoftwo
 \else \expandafter \@secondoftwo
 \fi
}%
\providecommand \natexlab [1]{#1}%
\providecommand \enquote  [1]{``#1''}%
\providecommand \bibnamefont  [1]{#1}%
\providecommand \bibfnamefont [1]{#1}%
\providecommand \citenamefont [1]{#1}%
\providecommand \href@noop [0]{\@secondoftwo}%
\providecommand \href [0]{\begingroup \@sanitize@url \@href}%
\providecommand \@href[1]{\@@startlink{#1}\@@href}%
\providecommand \@@href[1]{\endgroup#1\@@endlink}%
\providecommand \@sanitize@url [0]{\catcode `\\12\catcode `\$12\catcode
  `\&12\catcode `\#12\catcode `\^12\catcode `\_12\catcode `\%12\relax}%
\providecommand \@@startlink[1]{}%
\providecommand \@@endlink[0]{}%
\providecommand \url  [0]{\begingroup\@sanitize@url \@url }%
\providecommand \@url [1]{\endgroup\@href {#1}{\urlprefix }}%
\providecommand \urlprefix  [0]{URL }%
\providecommand \Eprint [0]{\href }%
\providecommand \doibase [0]{http://dx.doi.org/}%
\providecommand \selectlanguage [0]{\@gobble}%
\providecommand \bibinfo  [0]{\@secondoftwo}%
\providecommand \bibfield  [0]{\@secondoftwo}%
\providecommand \translation [1]{[#1]}%
\providecommand \BibitemOpen [0]{}%
\providecommand \bibitemStop [0]{}%
\providecommand \bibitemNoStop [0]{.\EOS\space}%
\providecommand \EOS [0]{\spacefactor3000\relax}%
\providecommand \BibitemShut  [1]{\csname bibitem#1\endcsname}%
\let\auto@bib@innerbib\@empty
\bibitem [{\citenamefont {Kroger}\ and\ \citenamefont
  {Reich}(1976)}]{KroRei76}%
  \BibitemOpen
  \bibfield  {author} {\bibinfo {author} {\bibfnamefont {L.~A.}\ \bibnamefont
  {Kroger}}\ and\ \bibinfo {author} {\bibfnamefont {C.~W.}\ \bibnamefont
  {Reich}},\ }\href@noop {} {\bibfield  {journal} {\bibinfo  {journal} {Nucl.
  Phys. A}\ }\textbf {\bibinfo {volume} {259}},\ \bibinfo {pages} {29}
  (\bibinfo {year} {1976})}\BibitemShut {NoStop}%
\bibitem [{\citenamefont {Helmer}\ and\ \citenamefont
  {Reich}(1994)}]{HelRei94}%
  \BibitemOpen
  \bibfield  {author} {\bibinfo {author} {\bibfnamefont {R.~G.}\ \bibnamefont
  {Helmer}}\ and\ \bibinfo {author} {\bibfnamefont {C.~W.}\ \bibnamefont
  {Reich}},\ }\href@noop {} {\bibfield  {journal} {\bibinfo  {journal} {Phys.
  Rev. C}\ }\textbf {\bibinfo {volume} {49}},\ \bibinfo {pages} {1845}
  (\bibinfo {year} {1994})}\BibitemShut {NoStop}%
\bibitem [{\citenamefont {Guimar\~aes{\rm -Filho}}\ and\ \citenamefont
  {Helene}(2005)}]{GuiHel05}%
  \BibitemOpen
  \bibfield  {author} {\bibinfo {author} {\bibfnamefont {Z.~O.}\ \bibnamefont
  {Guimar\~aes{\rm -Filho}}}\ and\ \bibinfo {author} {\bibfnamefont
  {O.}~\bibnamefont {Helene}},\ }\href@noop {} {\bibfield  {journal} {\bibinfo
  {journal} {Phys. Rev. C}\ }\textbf {\bibinfo {volume} {71}},\ \bibinfo
  {pages} {044303} (\bibinfo {year} {2005})}\BibitemShut {NoStop}%
\bibitem [{\citenamefont {Beck}\ \emph {et~al.}(2007)\citenamefont {Beck},
  \citenamefont {Becker}, \citenamefont {Beiersdorfer}, \citenamefont {Brown},
  \citenamefont {Moody}, \citenamefont {Wilhelmy}, \citenamefont {Porter},
  \citenamefont {Kilbourne},\ and\ \citenamefont {Kelley}}]{BecBecBei07}%
  \BibitemOpen
  \bibfield  {author} {\bibinfo {author} {\bibfnamefont {B.~R.}\ \bibnamefont
  {Beck}}, \bibinfo {author} {\bibfnamefont {J.~A.}\ \bibnamefont {Becker}},
  \bibinfo {author} {\bibfnamefont {P.}~\bibnamefont {Beiersdorfer}}, \bibinfo
  {author} {\bibfnamefont {G.~V.}\ \bibnamefont {Brown}}, \bibinfo {author}
  {\bibfnamefont {K.~J.}\ \bibnamefont {Moody}}, \bibinfo {author}
  {\bibfnamefont {J.~B.}\ \bibnamefont {Wilhelmy}}, \bibinfo {author}
  {\bibfnamefont {F.~S.}\ \bibnamefont {Porter}}, \bibinfo {author}
  {\bibfnamefont {C.~A.}\ \bibnamefont {Kilbourne}}, \ and\ \bibinfo {author}
  {\bibfnamefont {R.~L.}\ \bibnamefont {Kelley}},\ }\href@noop {} {\bibfield
  {journal} {\bibinfo  {journal} {Phys. Rev. Lett.}\ }\textbf {\bibinfo
  {volume} {98}},\ \bibinfo {pages} {142501} (\bibinfo {year}
  {2007})}\BibitemShut {NoStop}%
\bibitem [{Bec()}]{BecWuBei09}%
  \BibitemOpen
  \href@noop {} {}\bibinfo {note} {B. R. Beck, C. Y. Wu, P. Beiersdorfer, G. V.
  Brown, J. A. Becker, K. J. Moody, J. B. Wilhelmy, F. S. Porter, C. A.
  Kilbourne, and R. L. Kelley, Improved value for the energy splitting of the
  ground-state doublet in the nucleus $^{229m}$Th, Lawrence Livermore National
  Laboratory Technical Report No. LLNL-PROC-415170, 2009}\BibitemShut {NoStop}%
\bibitem [{\citenamefont {Seiferle}\ \emph {et~al.}(2019)\citenamefont
  {Seiferle}, \citenamefont {{\rm L. von der Wense}}, \citenamefont {Bilous},
  \citenamefont {Amersdorffer}, \citenamefont {Lemell}, \citenamefont
  {Libisch}, \citenamefont {Stellmer}, \citenamefont {Schumm}, \citenamefont
  {D\"{u}llmann}, \citenamefont {P\'{a}lffy},\ and\ \citenamefont
  {Thirolf}}]{SeiWenBil19}%
  \BibitemOpen
  \bibfield  {author} {\bibinfo {author} {\bibfnamefont {B.}~\bibnamefont
  {Seiferle}}, \bibinfo {author} {\bibnamefont {{\rm L. von der Wense}}},
  \bibinfo {author} {\bibfnamefont {P.~V.}\ \bibnamefont {Bilous}}, \bibinfo
  {author} {\bibfnamefont {I.}~\bibnamefont {Amersdorffer}}, \bibinfo {author}
  {\bibfnamefont {C.}~\bibnamefont {Lemell}}, \bibinfo {author} {\bibfnamefont
  {F.}~\bibnamefont {Libisch}}, \bibinfo {author} {\bibfnamefont
  {S.}~\bibnamefont {Stellmer}}, \bibinfo {author} {\bibfnamefont
  {T.}~\bibnamefont {Schumm}}, \bibinfo {author} {\bibfnamefont {C.~E.}\
  \bibnamefont {D\"{u}llmann}}, \bibinfo {author} {\bibfnamefont
  {A.}~\bibnamefont {P\'{a}lffy}}, \ and\ \bibinfo {author} {\bibfnamefont
  {P.~G.}\ \bibnamefont {Thirolf}},\ }\href@noop {} {\bibfield  {journal}
  {\bibinfo  {journal} {Nature (London)}\ }\textbf {\bibinfo {volume} {573}},\
  \bibinfo {pages} {243} (\bibinfo {year} {2019})}\BibitemShut {NoStop}%
\bibitem [{\citenamefont {Sikorsky}\ \emph {et~al.}(2020)\citenamefont
  {Sikorsky}, \citenamefont {Geist}, \citenamefont {Hengstler}, \citenamefont
  {Kempf}, \citenamefont {Gastaldo}, \citenamefont {Enss}, \citenamefont
  {Mokry}, \citenamefont {Runke}, \citenamefont {D\"ullmann}, \citenamefont
  {Wobrauschek}, \citenamefont {Beeks}, \citenamefont {Rosecker}, \citenamefont
  {Sterba}, \citenamefont {Kazakov}, \citenamefont {Schumm},\ and\
  \citenamefont {Fleischmann}}]{SikGeiHen20}%
  \BibitemOpen
  \bibfield  {author} {\bibinfo {author} {\bibfnamefont {T.}~\bibnamefont
  {Sikorsky}}, \bibinfo {author} {\bibfnamefont {J.}~\bibnamefont {Geist}},
  \bibinfo {author} {\bibfnamefont {D.}~\bibnamefont {Hengstler}}, \bibinfo
  {author} {\bibfnamefont {S.}~\bibnamefont {Kempf}}, \bibinfo {author}
  {\bibfnamefont {L.}~\bibnamefont {Gastaldo}}, \bibinfo {author}
  {\bibfnamefont {C.}~\bibnamefont {Enss}}, \bibinfo {author} {\bibfnamefont
  {C.}~\bibnamefont {Mokry}}, \bibinfo {author} {\bibfnamefont
  {J.}~\bibnamefont {Runke}}, \bibinfo {author} {\bibfnamefont {C.~E.}\
  \bibnamefont {D\"ullmann}}, \bibinfo {author} {\bibfnamefont
  {P.}~\bibnamefont {Wobrauschek}}, \bibinfo {author} {\bibfnamefont
  {K.}~\bibnamefont {Beeks}}, \bibinfo {author} {\bibfnamefont
  {V.}~\bibnamefont {Rosecker}}, \bibinfo {author} {\bibfnamefont {J.~H.}\
  \bibnamefont {Sterba}}, \bibinfo {author} {\bibfnamefont {G.}~\bibnamefont
  {Kazakov}}, \bibinfo {author} {\bibfnamefont {T.}~\bibnamefont {Schumm}}, \
  and\ \bibinfo {author} {\bibfnamefont {A.}~\bibnamefont {Fleischmann}},\
  }\href@noop {} {\bibfield  {journal} {\bibinfo  {journal} {Phys. Rev. Lett.}\
  }\textbf {\bibinfo {volume} {125}},\ \bibinfo {pages} {142503} (\bibinfo
  {year} {2020})}\BibitemShut {NoStop}%
\bibitem [{Pei()}]{PeiSchSaf21}%
  \BibitemOpen
  \href@noop {} {}\bibinfo {note} {E. Peik, T. Schumm, M. Safronova, A.
  P\'alffy, J. Weitenberg, and P.G. Thirolf (accepted to Quantum. Sci.
  Tech.)}\BibitemShut {NoStop}%
\bibitem [{\citenamefont {Peik}\ and\ \citenamefont {Tamm}(2003)}]{PeiTam03}%
  \BibitemOpen
  \bibfield  {author} {\bibinfo {author} {\bibfnamefont {E.}~\bibnamefont
  {Peik}}\ and\ \bibinfo {author} {\bibfnamefont {C.}~\bibnamefont {Tamm}},\
  }\href@noop {} {\bibfield  {journal} {\bibinfo  {journal} {Europhys. Lett.}\
  }\textbf {\bibinfo {volume} {61}},\ \bibinfo {pages} {181} (\bibinfo {year}
  {2003})}\BibitemShut {NoStop}%
\bibitem [{\citenamefont {{Flambaum}}(2006)}]{Fla06}%
  \BibitemOpen
  \bibfield  {author} {\bibinfo {author} {\bibfnamefont {V.~V.}\ \bibnamefont
  {{Flambaum}}},\ }\href {\doibase 10.1103/PhysRevLett.97.092502} {\bibfield
  {journal} {\bibinfo  {journal} {Phys. Rev. Lett.}\ }\textbf {\bibinfo
  {volume} {97}},\ \bibinfo {eid} {092502} (\bibinfo {year}
  {2006})}\BibitemShut {NoStop}%
\bibitem [{\citenamefont {Kozlov}\ \emph {et~al.}(2018)\citenamefont {Kozlov},
  \citenamefont {Safronova}, \citenamefont {{Crespo L\'{o}pez-Urrutia}},\ and\
  \citenamefont {Schmidt}}]{KozSafCre18}%
  \BibitemOpen
  \bibfield  {author} {\bibinfo {author} {\bibfnamefont {M.~G.}\ \bibnamefont
  {Kozlov}}, \bibinfo {author} {\bibfnamefont {M.~S.}\ \bibnamefont
  {Safronova}}, \bibinfo {author} {\bibfnamefont {J.~R.}\ \bibnamefont {{Crespo
  L\'{o}pez-Urrutia}}}, \ and\ \bibinfo {author} {\bibfnamefont {P.~O.}\
  \bibnamefont {Schmidt}},\ }\href@noop {} {\bibfield  {journal} {\bibinfo
  {journal} {Rev. Mod. Phys.}\ }\textbf {\bibinfo {volume} {90}},\ \bibinfo
  {pages} {045005} (\bibinfo {year} {2018})}\BibitemShut {NoStop}%
\bibitem [{\citenamefont {Bilous}\ \emph {et~al.}(2020)\citenamefont {Bilous},
  \citenamefont {Bekker}, \citenamefont {Berengut}, \citenamefont {Seiferle},
  \citenamefont {von~der Wense}, \citenamefont {Thirolf}, \citenamefont
  {Pfeifer}, \citenamefont {L\'opez-Urrutia},\ and\ \citenamefont
  {P\'alffy}}]{BilBekBer20}%
  \BibitemOpen
  \bibfield  {author} {\bibinfo {author} {\bibfnamefont {P.~V.}\ \bibnamefont
  {Bilous}}, \bibinfo {author} {\bibfnamefont {H.}~\bibnamefont {Bekker}},
  \bibinfo {author} {\bibfnamefont {J.~C.}\ \bibnamefont {Berengut}}, \bibinfo
  {author} {\bibfnamefont {B.}~\bibnamefont {Seiferle}}, \bibinfo {author}
  {\bibfnamefont {L.}~\bibnamefont {von~der Wense}}, \bibinfo {author}
  {\bibfnamefont {P.~G.}\ \bibnamefont {Thirolf}}, \bibinfo {author}
  {\bibfnamefont {T.}~\bibnamefont {Pfeifer}}, \bibinfo {author} {\bibfnamefont
  {J.~R.~C.}\ \bibnamefont {L\'opez-Urrutia}}, \ and\ \bibinfo {author}
  {\bibfnamefont {A.}~\bibnamefont {P\'alffy}},\ }\href@noop {} {\bibfield
  {journal} {\bibinfo  {journal} {Phys. Rev. Lett.}\ }\textbf {\bibinfo
  {volume} {124}},\ \bibinfo {pages} {192502} (\bibinfo {year}
  {2020})}\BibitemShut {NoStop}%
\bibitem [{\citenamefont {Cheung}\ \emph {et~al.}(2021)\citenamefont {Cheung},
  \citenamefont {Safronova},\ and\ \citenamefont {Porsev}}]{CheSafPor21}%
  \BibitemOpen
  \bibfield  {author} {\bibinfo {author} {\bibfnamefont {C.}~\bibnamefont
  {Cheung}}, \bibinfo {author} {\bibfnamefont {M.~S.}\ \bibnamefont
  {Safronova}}, \ and\ \bibinfo {author} {\bibfnamefont {S.~G.}\ \bibnamefont
  {Porsev}},\ }\href@noop {} {\bibfield  {journal} {\bibinfo  {journal}
  {Symmetry}\ }\textbf {\bibinfo {volume} {13}},\ \bibinfo {pages} {621}
  (\bibinfo {year} {2021})}\BibitemShut {NoStop}%
\bibitem [{\citenamefont {Porsev}\ \emph {et~al.}(2010)\citenamefont {Porsev},
  \citenamefont {Flambaum}, \citenamefont {Peik},\ and\ \citenamefont
  {Tamm}}]{PorFlaPei10}%
  \BibitemOpen
  \bibfield  {author} {\bibinfo {author} {\bibfnamefont {S.~G.}\ \bibnamefont
  {Porsev}}, \bibinfo {author} {\bibfnamefont {V.~V.}\ \bibnamefont
  {Flambaum}}, \bibinfo {author} {\bibfnamefont {E.}~\bibnamefont {Peik}}, \
  and\ \bibinfo {author} {\bibfnamefont {C.}~\bibnamefont {Tamm}},\ }\href@noop
  {} {\bibfield  {journal} {\bibinfo  {journal} {Phys. Rev. Lett.}\ }\textbf
  {\bibinfo {volume} {105}},\ \bibinfo {pages} {182501} (\bibinfo {year}
  {2010})}\BibitemShut {NoStop}%
\bibitem [{\citenamefont {Porsev}\ and\ \citenamefont
  {Flambaum}(2010{\natexlab{a}})}]{PorFla10ThIV}%
  \BibitemOpen
  \bibfield  {author} {\bibinfo {author} {\bibfnamefont {S.~G.}\ \bibnamefont
  {Porsev}}\ and\ \bibinfo {author} {\bibfnamefont {V.~V.}\ \bibnamefont
  {Flambaum}},\ }\href@noop {} {\bibfield  {journal} {\bibinfo  {journal}
  {Phys. Rev. A}\ }\textbf {\bibinfo {volume} {81}},\ \bibinfo {pages} {032504}
  (\bibinfo {year} {2010}{\natexlab{a}})}\BibitemShut {NoStop}%
\bibitem [{\citenamefont {Porsev}\ and\ \citenamefont
  {Flambaum}(2010{\natexlab{b}})}]{PorFla10ThII}%
  \BibitemOpen
  \bibfield  {author} {\bibinfo {author} {\bibfnamefont {S.~G.}\ \bibnamefont
  {Porsev}}\ and\ \bibinfo {author} {\bibfnamefont {V.~V.}\ \bibnamefont
  {Flambaum}},\ }\href@noop {} {\bibfield  {journal} {\bibinfo  {journal}
  {Phys. Rev. A}\ }\textbf {\bibinfo {volume} {81}},\ \bibinfo {pages} {042516}
  (\bibinfo {year} {2010}{\natexlab{b}})}\BibitemShut {NoStop}%
\bibitem [{\citenamefont {Sobelman}(1979)}]{Sob79}%
  \BibitemOpen
  \bibfield  {author} {\bibinfo {author} {\bibfnamefont {I.~I.}\ \bibnamefont
  {Sobelman}},\ }\href@noop {} {\emph {\bibinfo {title} {Atomic Spectra and
  Radiative Transitions}}}\ (\bibinfo  {publisher} {Springer-Verlag},\ \bibinfo
  {address} {Berlin},\ \bibinfo {year} {1979})\BibitemShut {NoStop}%
\bibitem [{\citenamefont {Varshalovich}\ \emph {et~al.}(1988)\citenamefont
  {Varshalovich}, \citenamefont {Moskalev},\ and\ \citenamefont
  {Khersonskii}}]{VarMosKhe88}%
  \BibitemOpen
  \bibfield  {author} {\bibinfo {author} {\bibfnamefont {D.~A.}\ \bibnamefont
  {Varshalovich}}, \bibinfo {author} {\bibfnamefont {A.~N.}\ \bibnamefont
  {Moskalev}}, \ and\ \bibinfo {author} {\bibfnamefont {V.~K.}\ \bibnamefont
  {Khersonskii}},\ }\href@noop {} {\emph {\bibinfo {title} {Quantum Theory of
  Angular Momentum}}}\ (\bibinfo  {publisher} {World Scientific},\ \bibinfo
  {address} {Singapore},\ \bibinfo {year} {1988})\BibitemShut {NoStop}%
\bibitem [{\citenamefont {Kozlov}\ \emph {et~al.}(2015)\citenamefont {Kozlov},
  \citenamefont {Porsev}, \citenamefont {Safronova},\ and\ \citenamefont
  {Tupitsyn}}]{KozPorSaf15}%
  \BibitemOpen
  \bibfield  {author} {\bibinfo {author} {\bibfnamefont {M.~G.}\ \bibnamefont
  {Kozlov}}, \bibinfo {author} {\bibfnamefont {S.~G.}\ \bibnamefont {Porsev}},
  \bibinfo {author} {\bibfnamefont {M.~S.}\ \bibnamefont {Safronova}}, \ and\
  \bibinfo {author} {\bibfnamefont {I.~I.}\ \bibnamefont {Tupitsyn}},\
  }\href@noop {} {\bibfield  {journal} {\bibinfo  {journal} {Comp. Phys.
  Comm.}\ }\textbf {\bibinfo {volume} {195}},\ \bibinfo {pages} {199} (\bibinfo
  {year} {2015})}\BibitemShut {NoStop}%
\bibitem [{\citenamefont {Kozlov}\ \emph {et~al.}(1996)\citenamefont {Kozlov},
  \citenamefont {Porsev},\ and\ \citenamefont {Flambaum}}]{KozPorFla96}%
  \BibitemOpen
  \bibfield  {author} {\bibinfo {author} {\bibfnamefont {M.~G.}\ \bibnamefont
  {Kozlov}}, \bibinfo {author} {\bibfnamefont {S.~G.}\ \bibnamefont {Porsev}},
  \ and\ \bibinfo {author} {\bibfnamefont {V.~V.}\ \bibnamefont {Flambaum}},\
  }\href@noop {} {\bibfield  {journal} {\bibinfo  {journal} {J. \ Phys. \ B}\
  }\textbf {\bibinfo {volume} {29}},\ \bibinfo {pages} {689} (\bibinfo {year}
  {1996})}\BibitemShut {NoStop}%
\bibitem [{\citenamefont {{Dzuba}}\ \emph {et~al.}(1996)\citenamefont
  {{Dzuba}}, \citenamefont {{Flambaum}},\ and\ \citenamefont
  {{Kozlov}}}]{DzuFlaKoz96}%
  \BibitemOpen
  \bibfield  {author} {\bibinfo {author} {\bibfnamefont {V.~A.}\ \bibnamefont
  {{Dzuba}}}, \bibinfo {author} {\bibfnamefont {V.~V.}\ \bibnamefont
  {{Flambaum}}}, \ and\ \bibinfo {author} {\bibfnamefont {M.~G.}\ \bibnamefont
  {{Kozlov}}},\ }\href {\doibase 10.1103/PhysRevA.54.3948} {\bibfield
  {journal} {\bibinfo  {journal} {\pra}\ }\textbf {\bibinfo {volume} {54}},\
  \bibinfo {pages} {3948} (\bibinfo {year} {1996})}\BibitemShut {NoStop}%
\bibitem [{\citenamefont {{Safronova}}\ \emph {et~al.}(2009)\citenamefont
  {{Safronova}}, \citenamefont {{Kozlov}}, \citenamefont {{Johnson}},\ and\
  \citenamefont {{Jiang}}}]{SafKozJoh09}%
  \BibitemOpen
  \bibfield  {author} {\bibinfo {author} {\bibfnamefont {M.~S.}\ \bibnamefont
  {{Safronova}}}, \bibinfo {author} {\bibfnamefont {M.~G.}\ \bibnamefont
  {{Kozlov}}}, \bibinfo {author} {\bibfnamefont {W.~R.}\ \bibnamefont
  {{Johnson}}}, \ and\ \bibinfo {author} {\bibfnamefont {D.}~\bibnamefont
  {{Jiang}}},\ }\href {\doibase 10.1103/PhysRevA.80.012516} {\bibfield
  {journal} {\bibinfo  {journal} {\pra}\ }\textbf {\bibinfo {volume} {80}},\
  \bibinfo {eid} {012516} (\bibinfo {year} {2009})}\BibitemShut {NoStop}%
\bibitem [{\citenamefont {Dykhne}\ and\ \citenamefont
  {Tkalya}(1998)}]{DykTka98}%
  \BibitemOpen
  \bibfield  {author} {\bibinfo {author} {\bibfnamefont {A.~M.}\ \bibnamefont
  {Dykhne}}\ and\ \bibinfo {author} {\bibfnamefont {E.~V.}\ \bibnamefont
  {Tkalya}},\ }\href@noop {} {\bibfield  {journal} {\bibinfo  {journal} {Pis'ma
  Zh. Eksp. Teor. Fiz.}\ }\textbf {\bibinfo {volume} {67}},\ \bibinfo {pages}
  {233} (\bibinfo {year} {1998})},\ \bibinfo {note} {[JETP \ Lett. {\bf 67},
  251 (1998)]}\BibitemShut {NoStop}%
\bibitem [{\citenamefont {Ruchowska}\ \emph {et~al.}(2006)\citenamefont
  {Ruchowska}, \citenamefont {P\l{}\'ociennik}, \citenamefont
  {\ifmmode~\dot{Z}\else \.{Z}\fi{}ylicz}, \citenamefont {Mach}, \citenamefont
  {Kvasil}, \citenamefont {Algora}, \citenamefont {Amzal}, \citenamefont
  {B\"ack}, \citenamefont {Borge}, \citenamefont {Boutami}, \citenamefont
  {Butler}, \citenamefont {Cederk\"all}, \citenamefont {Cederwall},
  \citenamefont {Fogelberg}, \citenamefont {Fraile}, \citenamefont {Fynbo},
  \citenamefont {Hageb\o{}}, \citenamefont {Hoff}, \citenamefont {Gausemel},
  \citenamefont {Jungclaus}, \citenamefont {Kaczarowski}, \citenamefont
  {Kerek}, \citenamefont {Kurcewicz}, \citenamefont {Lagergren}, \citenamefont
  {Nacher}, \citenamefont {Rubio}, \citenamefont {Syntfeld}, \citenamefont
  {Tengblad}, \citenamefont {Wasilewski},\ and\ \citenamefont
  {Weissman}}]{RucPloZyl06}%
  \BibitemOpen
  \bibfield  {author} {\bibinfo {author} {\bibfnamefont {E.}~\bibnamefont
  {Ruchowska}}, \bibinfo {author} {\bibfnamefont {W.~A.}\ \bibnamefont
  {P\l{}\'ociennik}}, \bibinfo {author} {\bibfnamefont {J.}~\bibnamefont
  {\ifmmode~\dot{Z}\else \.{Z}\fi{}ylicz}}, \bibinfo {author} {\bibfnamefont
  {H.}~\bibnamefont {Mach}}, \bibinfo {author} {\bibfnamefont {J.}~\bibnamefont
  {Kvasil}}, \bibinfo {author} {\bibfnamefont {A.}~\bibnamefont {Algora}},
  \bibinfo {author} {\bibfnamefont {N.}~\bibnamefont {Amzal}}, \bibinfo
  {author} {\bibfnamefont {T.}~\bibnamefont {B\"ack}}, \bibinfo {author}
  {\bibfnamefont {M.~G.}\ \bibnamefont {Borge}}, \bibinfo {author}
  {\bibfnamefont {R.}~\bibnamefont {Boutami}}, \bibinfo {author} {\bibfnamefont
  {P.~A.}\ \bibnamefont {Butler}}, \bibinfo {author} {\bibfnamefont
  {J.}~\bibnamefont {Cederk\"all}}, \bibinfo {author} {\bibfnamefont
  {B.}~\bibnamefont {Cederwall}}, \bibinfo {author} {\bibfnamefont
  {B.}~\bibnamefont {Fogelberg}}, \bibinfo {author} {\bibfnamefont {L.~M.}\
  \bibnamefont {Fraile}}, \bibinfo {author} {\bibfnamefont {H.~O.~U.}\
  \bibnamefont {Fynbo}}, \bibinfo {author} {\bibfnamefont {E.}~\bibnamefont
  {Hageb\o{}}}, \bibinfo {author} {\bibfnamefont {P.}~\bibnamefont {Hoff}},
  \bibinfo {author} {\bibfnamefont {H.}~\bibnamefont {Gausemel}}, \bibinfo
  {author} {\bibfnamefont {A.}~\bibnamefont {Jungclaus}}, \bibinfo {author}
  {\bibfnamefont {R.}~\bibnamefont {Kaczarowski}}, \bibinfo {author}
  {\bibfnamefont {A.}~\bibnamefont {Kerek}}, \bibinfo {author} {\bibfnamefont
  {W.}~\bibnamefont {Kurcewicz}}, \bibinfo {author} {\bibfnamefont
  {K.}~\bibnamefont {Lagergren}}, \bibinfo {author} {\bibfnamefont
  {E.}~\bibnamefont {Nacher}}, \bibinfo {author} {\bibfnamefont
  {B.}~\bibnamefont {Rubio}}, \bibinfo {author} {\bibfnamefont
  {A.}~\bibnamefont {Syntfeld}}, \bibinfo {author} {\bibfnamefont
  {O.}~\bibnamefont {Tengblad}}, \bibinfo {author} {\bibfnamefont {A.~A.}\
  \bibnamefont {Wasilewski}}, \ and\ \bibinfo {author} {\bibfnamefont
  {L.}~\bibnamefont {Weissman}},\ }\href {\doibase 10.1103/PhysRevC.73.044326}
  {\bibfield  {journal} {\bibinfo  {journal} {Phys. Rev. C}\ }\textbf {\bibinfo
  {volume} {73}},\ \bibinfo {pages} {044326} (\bibinfo {year}
  {2006})}\BibitemShut {NoStop}%
\bibitem [{\citenamefont {Minkov}\ and\ \citenamefont
  {P\'alffy}(2019)}]{MinPal19}%
  \BibitemOpen
  \bibfield  {author} {\bibinfo {author} {\bibfnamefont {N.}~\bibnamefont
  {Minkov}}\ and\ \bibinfo {author} {\bibfnamefont {A.}~\bibnamefont
  {P\'alffy}},\ }\href@noop {} {\bibfield  {journal} {\bibinfo  {journal}
  {Phys. Rev. Lett.}\ }\textbf {\bibinfo {volume} {122}},\ \bibinfo {pages}
  {162502} (\bibinfo {year} {2019})}\BibitemShut {NoStop}%
\bibitem [{\citenamefont {Strizhov}\ and\ \citenamefont
  {Tkalya}(1991)}]{StrTka91}%
  \BibitemOpen
  \bibfield  {author} {\bibinfo {author} {\bibfnamefont {V.~F.}\ \bibnamefont
  {Strizhov}}\ and\ \bibinfo {author} {\bibfnamefont {E.~V.}\ \bibnamefont
  {Tkalya}},\ }\href@noop {} {\bibfield  {journal} {\bibinfo  {journal} {Zh.
  Eksp. Teor. Fiz.}\ }\textbf {\bibinfo {volume} {99}},\ \bibinfo {pages} {697}
  (\bibinfo {year} {1991})},\ \bibinfo {note} {[Sov. Phys.-JETP {\bf 72}, 387
  (1991)]}\BibitemShut {NoStop}%
\bibitem [{Bem()}]{BemMcGPor88}%
  \BibitemOpen
  \href@noop {} {}\bibinfo {note} {C.~E.~Bemis, Jr., P.~K.~McGowan,
  F.~C.~Porter {\it et al.,} Phys. Scr. {\bf 38}, 657 (1988)}\BibitemShut
  {NoStop}%
\end{thebibliography}

%

\end{document}